\documentclass[runningheads]{llncs}
\usepackage[T1]{fontenc}
\usepackage{graphicx}
\usepackage[T1]{fontenc}
\usepackage{comment}
\usepackage[utf8]{inputenc}
\usepackage{tipa}
\usepackage{array}
\usepackage{amsmath}
\usepackage{multirow}
\usepackage{balance}
\begin{document}
\title{Fairness Evaluation of Edge-AI Implementation for Cleft Lip and Palate Speech ASR}
%
%
\author{Susmita Bhattacharjee\inst{1}\orcidID{0009-0009-0169-3420} \and
Himashri Deka\inst{1}\orcidID{0009-0003-5705-9468} \and
H.S. Shekhawat\inst{1}\orcidID{0000-0002-5174-8903} \and
S.R.M. Prasanna\inst{2}\orcidID{0000-0002-8135-7938}}
\authorrunning{S. Bhattacharjee et al.}
\titlerunning{Fairness Evaluation of Edge-AI Implementation for CLP speech ASR}
%
\institute{Indian Institute of Technology Guwahati, India \\
\email{\{sbhattacharjee,himashri.deka,h.s.shekhawat\}@iitg.ac.in}\\
\and
Indian Institute of Technology Dharwad, India\\
\email{prasanna@iitdh.ac.in}}
\maketitle              
\begin{abstract}

Automatic speech recognition (ASR) remains challenging for individuals with cleft lip and palate (CLP) because of limited pathological speech data and large variations in speech characteristics across speakers and severity levels. These recognition difficulties can reduce the accessibility of voice-based human–computer interaction, particularly when cloud-based ASR services are unavailable or unreliable. This work investigates a severity-aware and edge-deployable ASR framework for improving recognition of CLP speech using Whisper-small. The model was fine-tuned using different combinations of normal and CLP speech representing mild, moderate, and severe conditions, together with a CLP-only training configuration, to examine how the inclusion of different severity levels influences recognition performance and fairness across speakers. The pretrained model produced pooled word error rate (WER) and phoneme error rate (PER) values of 62.46\% and 52.72\%, respectively. Severity-aware fine-tuning substantially improved performance, reducing the best pooled WER to 22.72\% and the best pooled PER to 18.44\%. Training with a broader representation of CLP severity levels also provided the best overall balance between recognition accuracy and performance consistency across severity groups. Deployment on an NVIDIA Jetson platform demonstrated real-time inference for all fine-tuned models, with real-time factors of 0.167–0.171 and peak GPU memory usage of approximately 566 MB. The results demonstrate that incorporating severity diversity during ASR adaptation can substantially improve recognition of CLP speech while reducing performance disparities across severity groups. The proposed approach further enables low-latency, Internet-independent speech interaction on edge devices, supporting more accessible and inclusive voice-based human–computer interaction for individuals with CLP.

\end{abstract}
\section{Introduction}
Cleft Lip and Palate (CLP) condition can cause normal speech to become atypical by introducing hypernasality, articulatory issues and weakening of plosives~\cite{evaluationand,methofperceptualassessment,clpnatureandremediation}. However, recognition performance must remain effective across CLP severity levels.
Recently, for objective assessment of CLP speech, ASR has been used, with recognition performance shown to correlate with intelligibility~\cite{scipioni09_interspeech,maier08_interspeech}. However, heterogeneity due to severity-dependent acoustic and articulatory variability continues to defy conventional and pretrained ASR systems~\cite{bhattacharjee2026improvingasrfairnesscleft}. The opening between the lip and nasal cavity results in speech that is breathy and highly nasalized~\cite{Zajac2011ReliabilityAV,Whitehill2004SinglewordII}.
It has been reported that ASR systems exhibit significant performance gap across user groups. 
The disparities in performance are associated with demographic and linguistic factors with phoneme-level analysis revealing underlying bias in the recognition~\cite{feng2021quantifyingbiasautomaticspeech}. 
Fair-Speech analysis has been found to provide reliable emphasis on the subgroup-level evaluation rather than depending only on aggregate WER~\cite{veliche24_interspeech}. However this study in disparity has not been explored much in the pathological speech domain. Here in majority of the cases, this impairment severity itself act as an important source of disparity. 
Recently transformer-based ASR models (such as wav2vec2, whisper) have been found to provide reduced WER as compared to traditional GMM-HMM and other deep learning models~\cite{xlsr,xlsr0,GMM,whisper}.
Whisper, proposed by Radford et al.~\cite{whisper} is trained on around 680,000 hours of multilingual and multitask speech datasets. It uses an encoder-decoder approach and conducts end-to-end transcription of speech in which raw speech is converted into log-Mel spectrograms, and then text is generated. Due to being trained on such a vast dataset, Whisper has demonstrated robust zero-shot recognition across different accents, acoustic conditions and domains~\cite{whisper}.
However, given the extreme heterogeneity in CLP speech, Whisper also underperforms in most of the cases. In one of our reported works in CLP ASR using Whisper and other foundation models, to mitigate the disparity in the available ASR models, severity-aware data mixing has been shown to improve both recognition performance and fairness across the categories of CLP- Normal, Mild, Moderate and Severe speech~\cite{bhattacharjee2026improvingasrfairnesscleft}.
Also, to further complement WER, phoneme-level analysis has been found to contribute quite significantly by identifying specific recognition confusions~\cite{feng2021quantifyingbiasautomaticspeech}.
Another reported contribution to edge computing focused on practical deployment in resource-constrained devices by reducing dependence on cloud infrastructure~\cite{electronics10212697}. 

\begin{figure*}[t]
	\centering	
    \includegraphics[width=\textwidth]{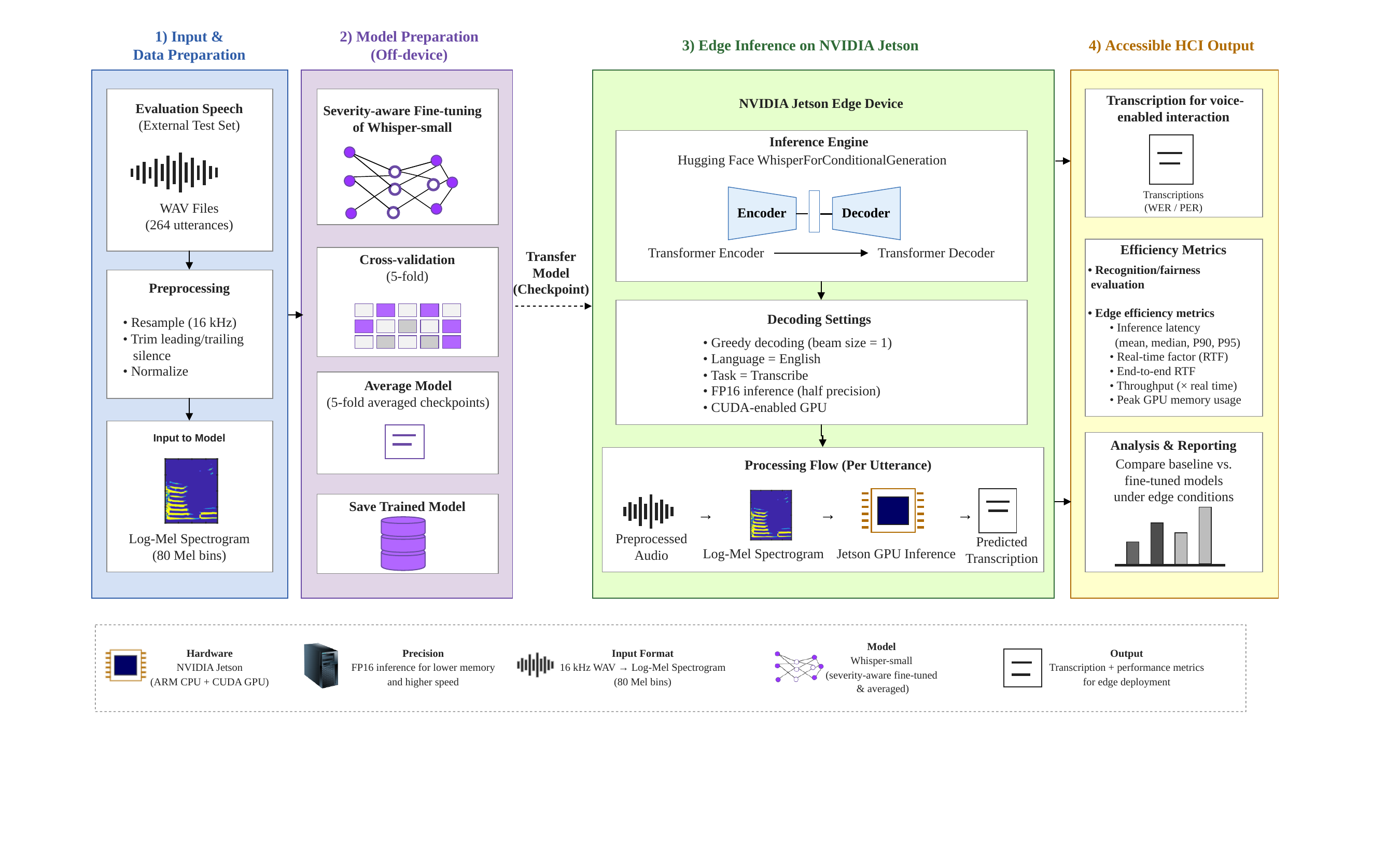}
	\caption{Generalized block diagram of the Jetson-based edge-ASR pipeline. Severity-aware Whisper-small models are fine-tuned off-device, averaged across five folds, and deployed on an NVIDIA Jetson for FP16 GPU inference. Preprocessed speech is converted to log-Mel spectrograms and decoded into transcriptions, while WER/PER, fairness, latency, RTF, throughput, and GPU-memory metrics are reported to assess both recognition performance and suitability for accessible CLP-oriented human–computer interaction.}
	\label{fig1}
 \vspace{-0.5 cm}
\end{figure*}

Edge-AI is particularly relevant to CLP ASR because continuous Internet connectivity cannot always be assumed. On-device processing reduces dependence on cloud infrastructure, enables low-latency interaction, and limits the need to transmit speech data externally.
Local computation allows the ASR to be available without depending on cloud computing.
Additionally, processing the speech locally leads to fast response times, reduces the transfer of speech data, and hence ensures practical and privacy-conscious assistive communication.
However, effective edge deployment is not equally fair across all CLP severity levels.
In this work, Whisper-small\footnote{\url{https://huggingface.co/openai/whisper-small}} is used. Several severity-aware Whisper-small models were fine-tuned using different combinations of Normal, Mild, Moderate, and Severe CLP speech. The resulting trained models were saved, transferred to an NVIDIA Jetson edge platform, deployed on the device, and evaluated through direct on-device inference on the evaluation speech. This enabled both recognition accuracy and computational performance to be assessed under practical edge-deployment conditions as seen from Figure~\ref{fig1}.
From an Intelligent Human–Computer Interaction perspective, the proposed system represents an adaptive and accessible speech interface designed for users whose speech characteristics differ substantially from those represented in conventional ASR systems. The intelligence of the interaction lies not only in converting speech to text, but in adapting the recognition model to severity-dependent CLP speech, reducing recognition disparity across user groups, and enabling responsive on-device interaction under resource and connectivity constraints. Edge deployment further supports practical assistive HCI by providing low-latency and reduced cloud dependence speech interaction without requiring continuous access to cloud-based services.
Therefore, this research aims at investigating the severity-wise fairness of an Edge-AI-based Whisper-small ASR model for CLP speech. This is done through comparing pretrained and severity-aware fine-tuned models based on severity-wise WER, phoneme-wise error rate, latency, real-time factor, throughput and memory usage.

\section{Dataset used}

\textbf{NMCPC dataset:} The NMCPC dataset contains speech from 41 CLP speakers (22 male, 19 female) and 24 Normal speakers (20 male, 4 female), aged 9–13 years, with CLP speech labelled as Mild, Moderate, or Severe [16]. Speaker-disjoint training, development, and evaluation partitions were used. To control training-set size, each configuration contained 280 training utterances: NO (280 Normal), NOMI (140+140), NOMIMO (93+93+94), NOMIMOSE (70 per group), and CLP-only (93+93+94). The evaluation set contained 264 utterances, with 66 from each severity group.

\subsection{Fairness as a Metric}\label{ob_study2}

Following Howard et al.~\cite{fairness1} and Liang et al.~\cite{fairness2}, fairness is evaluated using a Fairness Score (FS) that jointly considers the average error rate and the error disparity between Normal ($G_1$) and CLP ($G_2$) speech:

\begin{equation}
FS = -\alpha \cdot \text{Average Error Rate}
-\beta \cdot \text{Error Disparity},
\quad \alpha,\beta \geq 0
\label{eq:fairness_score}
\end{equation}

where

\begin{equation}
\text{Average Error Rate}
=
\frac{\text{error}(G_1)+\text{error}(G_2)}{2},
\label{eq:aer}
\end{equation}

and

\begin{equation}
\text{Error Disparity}
=
\left|\text{error}(G_1)-\text{error}(G_2)\right|.
\label{eq:ed}
\end{equation}

Here, $\text{error}(G_1)$ and $\text{error}(G_2)$ denote the WERs of Normal and CLP speech, respectively, while $\alpha$ and $\beta$ control the relative importance of overall error and disparity. The score lies in the range $-\infty \leq FS \leq 0$, with values closer to zero indicating better overall performance and greater fairness. A lower average error rate reflects better recognition accuracy, whereas a smaller error disparity indicates more balanced performance between Normal and CLP speech. Since CLP datasets have unequal number of files for all severity categories. $WER_{pooled}$ has been considered instead of averaged error.

\section{Experimental Setup} 

The inference efficiency of the ASR systems was evaluated on an NVIDIA Jetson platform using identical evaluation utterances and decoding settings. The pretrained Whisper-small model served as a zero-shot baseline, while the proposed systems used five-fold averaged models. Inference was performed with Hugging Face WhisperForConditionalGeneration on the CUDA-enabled GPU using FP16 precision, greedy decoding (beam size 1), English transcription, silence trimming, and utterance-wise processing. Computational performance was assessed using inference latency, real-time factor (RTF), end-to-end RTF, throughput and peak GPU memory. Latency was reported using mean, median and P95 statistics to account for occasional long autoregressive decoding sequences, while model loading time was measured separately from steady-state inference.




\begin{table*}[t]
\centering
\caption{Evaluation metrics used for ASR recognition and deployment analysis.}
\label{tab:evaluation_metrics}

\scriptsize
\renewcommand{\arraystretch}{1.35}
\setlength{\tabcolsep}{4pt}

\begin{tabular}{
|p{2.0cm}|
p{4.0cm}|
p{5cm}|
p{1.2cm}|
}
\hline

\textbf{Metric} &
\textbf{Formula} &
\textbf{Interpretation} &
\textbf{Better} \\
\hline

Word Error Rate (WER) &
$\displaystyle
\mathrm{WER}=
\frac{S+D+I}{N}\times100
$ &
Percentage of word-level recognition errors in the predicted transcription. &
Lower \\
\hline

Pooled WER &
$\displaystyle
\mathrm{WER}_{\mathrm{pooled}}=
\frac{\sum(S+D+I)}{\sum N}\times100
$ &
Corpus-level WER calculated jointly over all evaluated severity groups. &
Lower \\
\hline




End-to-end latency &
$\displaystyle
L_{\mathrm{E2E}}=
L_{\mathrm{pre}}+
L_{\mathrm{inf}}+
L_{\mathrm{dec}}
$ &
Total ASR processing time including preprocessing, model inference, and decoding. &
Lower \\
\hline

Real-Time Factor (RTF) &
$\displaystyle
\mathrm{RTF}=
\frac{T_{\mathrm{inf}}}{T_{\mathrm{aud}}}
$ &
Ratio of inference time to input-audio duration. An RTF below 1 indicates faster-than-real-time processing. &
Lower \\
\hline

End-to-end RTF &
$\displaystyle
\mathrm{RTF}_{\mathrm{E2E}}=
\frac{T_{\mathrm{E2E}}}{T_{\mathrm{aud}}}
$ &
Real-time factor considering the complete ASR pipeline rather than model inference alone. &
Lower \\
\hline

Real-time throughput &
$\displaystyle
\mathrm{Throughput}_{\times\mathrm{RT}}=
\frac{1}{\mathrm{RTF}}
$ &
Number of times faster than real time the system processes the speech signal. &
Higher \\
\hline

Utterance throughput &
$\displaystyle
\mathrm{Utt./s}=
\frac{M}{T_{\mathrm{inf,total}}}
$ &
Number of utterances processed per second of cumulative inference time. &
Higher \\
\hline


Peak GPU memory &
$\displaystyle
M_{\mathrm{GPU}}^{\mathrm{peak}}=
\max_t M_{\mathrm{GPU}}(t)
$ &
Maximum GPU memory allocated during inference. &
Lower \\
\hline

\end{tabular}
\end{table*}

\subsection{Deployment Results for Pretrained and finetuned Whisper-small}
The pretrained Whisper-small baseline processed all 264 evaluation utterances (451.45 s of speech) with mean and P95 inference latencies of 0.363 s and 0.385 s, respectively, an RTF of 0.212, and $4.72\times$ real-time throughput. Occasional repetitive decoder outputs and the first-utterance cold-start increased latency; after excluding these outliers, the steady-state RTF improved to 0.181 (5.52× real time).

\begin{table*}[ht]
\centering
\caption{Severity-wise recognition performance of the pretrained and fine-tuned Whisper-small models. Best results in each column are shown in bold.}
\label{tab:wer_per_results}

\scriptsize
\renewcommand{\arraystretch}{1.18}
\setlength{\tabcolsep}{3.0pt}

\begin{tabular}{
|p{2.7cm}|
>{\centering\arraybackslash}p{1.35cm}|
>{\centering\arraybackslash}p{1.35cm}|
>{\centering\arraybackslash}p{1.45cm}|
>{\centering\arraybackslash}p{1.35cm}|
>{\centering\arraybackslash}p{1.45cm}|
>{\centering\arraybackslash}p{1.45cm}|
}
\hline

\textbf{Model / Training} &
\textbf{Normal WER (\%)} &
\textbf{Mild WER (\%)} &
\textbf{Moderate WER (\%)} &
\textbf{Severe WER (\%)} &
\textbf{Pooled WER (\%)} &
\textbf{Pooled PER (\%)} \\
\hline

Direct Whisper-small
& 16.00 & 33.20 & 108.27 & 92.98 & 62.46 & 52.72 \\
\hline

Finetuned NO
& 4.80 & 16.21 & 42.13 & 80.58 & 35.54 & 29.44 \\
\hline

Finetuned NOMI
& 4.40 & 10.28 & 29.13 & 70.66 & 28.23 & 23.44 \\
\hline

Finetuned NOMIMO
& 4.80 & 7.91 & 20.47 & 59.50 & 22.82 & \textbf{18.44} \\
\hline

\textbf{Finetuned NOMIMOSE}
& 4.80 & 6.72 & 25.59 & 54.96
& \textbf{22.72} & 18.54 \\
\hline

Finetuned CLP
& 6.00 & 9.88 & 18.50 & 58.26 & 22.82 & 18.78 \\
\hline

\end{tabular}
\end{table*}
Model loading time ranged from 2.877 to 2.964 s across the evaluated configurations.
Median inference latency ranged from 0.274 to 0.301 s, with NOMIMOSE achieving the lowest value of 0.274 s. Peak GPU memory usage remained constant at 566.32 MB for all models. The corresponding utterance-processing throughput ranged from 2.76 utterances/s for the pretrained model to 3.50 utterances/s for NOMIMOSE which achieved the lowest pooled WER and fairness score as well. The $CLP-only$ model with the fairness score of $-22.85$ is taken the baseline for CLP since available LLMs are trained mainly with Normal speech and hence NOMIMOSE is the closest to it. This shows that minimizing aggregate recognition error and minimizing Normal–CLP performance disparity are related but distinct objectives. For assistive HCI, both measures are relevant because an interface should provide not only high average recognition accuracy but also reasonably balanced performance across impairment severities.

\begin{table*}[ht]
\centering
\caption{Fairness and edge-inference performance of the pretrained and fine-tuned Whisper-small models on NVIDIA Jetson. Higher FS and throughput, and lower latency and RTF values, indicate better performance. Best results in each column are shown in bold.}
\label{tab:edge_deployment_results}

\scriptsize
\renewcommand{\arraystretch}{1.18}
\setlength{\tabcolsep}{2.8pt}

\begin{tabular}{
|p{2.6cm}|
>{\centering\arraybackslash}p{1.45cm}|
>{\centering\arraybackslash}p{1.25cm}|
>{\centering\arraybackslash}p{1.25cm}|
>{\centering\arraybackslash}p{1.25cm}|
>{\centering\arraybackslash}p{1.10cm}|
>{\centering\arraybackslash}p{1.10cm}|
>{\centering\arraybackslash}p{1.15cm}|
}
\hline

\multirow{2}{*}{\textbf{Model}} &
\multirow{2}{*}{\textbf{FS}} &
\multicolumn{3}{c|}{\textbf{Latency (s)}} &
\multicolumn{2}{c|}{\textbf{Real-Time Factor}} &
\multirow{2}{*}{\textbf{$\times$RT}} \\
\cline{3-7}

&
&
\textbf{Mean} &
\textbf{P95} &
\textbf{E2E} &
\textbf{RTF} &
\textbf{E2E RTF} &
\\
\hline

Direct Whisper-small
& -62.30
& 0.363
& 0.385
& 0.508
& 0.212
& 0.297
& 4.72 \\
\hline

Finetuned NO
& -38.52
& 0.289
& 0.380
& 0.359
& 0.169
& 0.210
& 5.92 \\
\hline

Finetuned NOMI
& -30.26
& 0.287
& 0.366
& 0.358
& 0.168
& 0.209
& 5.95 \\
\hline

Finetuned NOMIMO
& -23.65
& 0.292
& 0.368
& 0.360
& 0.171
& 0.210
& 5.86 \\
\hline

\textbf{Finetuned NOMIMOSE}
& -23.50
& \textbf{0.286}
& 0.358
& \textbf{0.354}
& \textbf{0.167}
& \textbf{0.207}
& \textbf{5.99} \\
\hline

Finetuned CLP
& -22.85
& 0.287
& \textbf{0.356}
& 0.356
& 0.168
& 0.208
& 5.97 \\
\hline

\end{tabular}
\end{table*}

\subsection{WER Analysis and Phoneme Analysis}
For the Direct Whisper-small baseline, recurring phoneme substitutions included /b/$\rightarrow$/m/, /t/$\rightarrow$/k/, /k/$\rightarrow$/t/, /z/$\rightarrow$/t/, /v/$\rightarrow$/m/, /s/$\rightarrow$/\textipa{S}/, /z/$\rightarrow$/s/, and /\textipa{I}/$\leftrightarrow$/i/, with /b/$\rightarrow$/m/ occurring 26 times. These represent text-derived ASR phoneme confusions rather than direct acoustic or articulatory errors. Fine-tuning markedly reduced the pooled PER from 52.72\% for Direct Whisper-small to 29.44\% (NO), 23.44\% (NOMI), 18.44\% (NOMIMO), 18.54\% (NOMIMOSE), and 18.78\% (CLP). NOMIMO achieved the lowest pooled PER, while NOMIMOSE performed best for Mild (3.01\%) and Severe (47.32\%) speech, and CLP achieved the lowest Moderate PER (14.90\%).
\subsection{Edge Deployment performance}
The pretrained and fine-tuned Whisper-small models were evaluated under identical inference conditions using FP16 precision, a beam size of one, and GPU acceleration. All 264 evaluation utterances, corresponding to 451.45 s of processed speech, were successfully decoded by each system. The pretrained Whisper-small baseline achieved an overall RTF of 0.212, corresponding to $4.72\times$ real-time processing, with mean and P95 inference latencies of 0.363 s and 0.385 s, respectively. The fine-tuned models exhibited RTFs between 0.167 and 0.171 and processed speech at approximately 5.86–5.99× real time. Among these, NOMIMOSE achieved an RTF of 0.167, an end-to-end RTF of 0.207, and a mean inference latency of 0.286 s while obtaining the lowest pooled WER of 22.72\%. Its audio throughput reached 5.99× real time, with an inference throughput of 3.50 utterances/s. All models required approximately 566.32 MB of peak GPU memory, indicating that the recognition improvements did not introduce additional inference-memory overhead.
Importantly, the gains in recognition accuracy were obtained without increasing the underlying model architecture or its inference-memory requirement. Compared with the pretrained Whisper-small baseline, whose pooled WER was 62.46\%, the NOMIMOSE configuration reduced the pooled WER to 22.72\% while simultaneously reducing the RTF from 0.212 to 0.167. These results demonstrate that severity-aware fine-tuning can substantially improve recognition of CLP speech while retaining efficient real-time inference characteristics suitable for edge deployment.

\section{Conclusion}
This study evaluated severity-aware Whisper-small adaptation for CLP speech recognition on an NVIDIA Jetson platform. Fine-tuning reduced pooled WER from 62.46\% for pretrained Whisper to 22.72\% with NOMIMOSE, while NOMIMO achieved the lowest pooled PER of 18.44\% and fairness score almost similar to the CLP-only model highlighting that aggregate accuracy and Normal–CLP disparity are related but distinct objectives. Fine-tuned models achieved RTFs of 0.167--0.171 with approximately 566 MB peak GPU memory, demonstrating faster-than-real-time on-device inference without additional model-size overhead. Severe speech remained the most challenging condition. Overall, the results support severity-aware edge ASR as an accessible IHCI solution for CLP speech in settings where continuous cloud connectivity cannot be guaranteed.

\bibliographystyle{IEEEtran}
\bibliography{mybib}
\balance
\end{document}